\documentstyle[]{article}  

\begin{document}


\bibliographystyle{unsrt}    

\def\a{\alpha}\def\b{\beta}\def\g{\gamma}\def\d{\delta}\def\D{\Delta}
\def\e{\epsilon}\def\k{\kappa}\def\l{\lambda}\def\L{\Lambda}
\def\s{\sigma}\def\S{\Sigma}\def\Th{\Theta}\def\th{\theta}
\def\om{\omega}\def\Om{\Omega}\def\G{\Gamma}\def\v{\varepsilon}

\title{L\'evy Anomalous Diffusion \\ and Fractional Fokker--Planck Equation}
\date{(Submitted to Physica A, 12 June 1997, revised version: 22 January 1999, accepted: December 1999)}
\author{V.V. Yanovsky, A.V. Chechkin \\
Turbulence Research \\ Institute for Single Crystals\\
 National Acad. Sci. Ukraine, \\ Lenin ave. 60, Kharkov 310001, Ukraine\\
\and D. Schertzer \\ Laboratoire de Mod\'elisation en M\'ecanique,
Tour 66, Boite 162
\\ Universit\'e
Pierre et Marie Curie,\\ 4 Place Jussieu F-75252 Paris Cedex 05,
France.\\ \and A.V. Tur(Tour) \\ Centre d'Etude Spatiale des
Rayonnements
\\ 9 Avenue Colonel-Roche,BP 4346, 31028 Toulouse Cedex 4, France\\}


\maketitle

\begin{abstract}

We demonstrate that the Fokker-Planck equation can be generalized
 into a 'Fractional Fokker-Planck' equation, i.e. an equation which
includes fractional space differentiations, in order to encompass
the wide class of anomalous diffusions due to a L\'evy stable
stochastic forcing. A precise determination of this equation is
obtained by substituting a L\'evy stable source to the classical
gaussian one in the Langevin equation. This yields not only the
anomalous diffusion coefficient, but a non trivial fractional
operator which corresponds to the possible asymmetry of the L\'evy
stable source.  Both of them cannot be obtained by scaling
arguments. The (mono-) scaling behaviors of the Fractional
Fokker-Planck equation and of its solutions are analysed and a
generalization of the Einstein relation for the anomalous
diffusion coefficient is obtained.\\ This generalization yields a
straightforward  physical interpretation of the parameters of
L\'evy stable distributions. Furthermore, with the help of
important  examples, we show the applicability of the Fractional
Fokker-Planck equation in physics.
\\
\end{abstract}

\newpage%

\section{Introduction}\label{Introduction}

The Fokker--Planck equation is one of the classical, widely used
equations of statistical physics. It describes a broad spectrum of
problems related to the evolution of various dynamic systems under
the influence of stochastic forces and has numerous applications,
see, e.g. \cite{Kampen}. Usually, the Fokker--Planck equation can
be derived following the Langevin approach, that is, starting from
the stochastic "equation of motion" for the dynamic variable whose
probability distribution we are interested. In this approach, the
basic assumptions   on the "random force/source" in this equation
of motion are usually that they have: (i) Gaussian statistics, and
(ii) delta-correlated correlation. These two  assumptions yield
the (classical) Fokker--Planck equation.

These assumptions are physically motivated by the fact that the
random source is a sum of a large number of independent identical
random "pulses". If these quantities possess a finite variance,
then, according to the Central Limit Theorem, the distribution of
their sum tends to the normal law when the  number of pulses go to
infinity.

However, the Central Limit Theorem can be generalized for
independent identically distributed (i.i.d.) random variables
having non finite variance. Indeed, L\'evy and Khintchine
\cite{Khintchine, Levy,Feller,Zolotarev} discovered a broader
class of stable distributions. They correspond to the limit of
normalized sums of i.i.d. stochastic variables. Each stable law
has a characteristic index $\a ~~(0<\a \leq 2)$, often called the
L\'evy stability index or the L\'evy index, which is the critical
order for the convergence of statistical moments. Indeed, a
statistical moment of a given stable law is  finite only if its
order $\mu$ is strictly smaller than its L\'evy index  $\a$ (i.e.
$ \mu<\a$). Every moment of  order higher order (including: $\mu
=\a$) are infinite or, as often said, divergent. The only
exception is the normal distribution which corresponds to the
particular stable law which has its  L\'evy's index $\a =2$ and
the exceptional property that all its moments are finite.

The classical and rather academic example of the application of
the L\'evy stable laws is the Holtzmark distribution
\cite{Holtzmark,Chandrasekar} which is the distribution function
of the gravitational force created at a randomly chosen point by a
given system of stars. It is assumed that the system of stars is a
(statistically) homogeneous set of physical points which mutually
interact according to the gravitation law. Using dimensional
consideration it can be shown \cite{Feller} with a rather
straightforward scaling argument that this distribution
corresponds to a stable law with index
 $\a = 3/2 $. We will see (Sect.\ref{applications}) that the following
developments allow to generalize broadly this result.

Some other examples of application in physics of stable
distributions can be found in review papers \cite{Bouchaud,West}
and in references therein.

However, let us recall that presumably the most well known
application of L\'evy stable laws in physics corresponds to the
anomalous diffusion associated with a L\'evy motion, also often
called a 'L\'evy flight' \cite{Mandelbrot}. Indeed, one expects
that with a L\'evy stable forcing the cloud of particles will
spread much faster (for large times: $t>>1$) than for a brownian
motion. More precisely, we will confirm that the radius $r(t)$ of
the cloud, at time $t$, has the following scaling law:

\begin{equation}\label{0}
r(t) \sim t^{1/\a}
\end{equation}

\noindent the lower bound being reached for the normal diffusion
$(\a = 2)$, Figs.1 and 2 display illustrations for comparison.

This scaling relation (Eq.\ref{0}) is obviously incompatible with
the Fokker-Planck equation, unless one substitutes  a formal
$(\alpha /2)-th$ power of the Laplacian to the classical
Laplacian, therefore considers a Fractional Fokker-Planck
equation, as suggested by \cite{Zaslavsky}. Several authors,
following different approaches, considered generalization of the
Fokker-Planck equation in order to encompass the L\'evy anomalous
diffusion. On the one hand, particular cases of the Fractional
Fokker-Planck equation were obtained \cite{Fogedby, Compte,
Chaves}. However, this is not the only way to generalize the
Fokker-Planck equation in order to respect the anomalous scaling
relation of (Eq.\ref{0}. Indeed, based on Tsallis\cite{Tsallis}'s
generalization of statistical mechanics, a nonlinear Fokker-Planck
equation \cite{Plastino,Tsallis_Bukman} has been introduced and it
was demonstrated \cite{Tsallis_Bukman, Boland} that its solution
respects also Eq.\ref{0}. These approaches will be discussed and
compared to ours in Sect. \ref{sec.comparison}.

However, the interest in L\'evy laws is not limited to anomalous
diffusion. For instance, the rather large subclass of extremal
$1/f$ or "pink" L\'evy noises have been attracting much attention
in the framework of multifractal fields. Indeed, it was shown that
they correspond \cite{Schertzer} to the attractive generators of
"universal multifractals", which are the limit processes, under
rather general conditions, of nonlinearly interacting i.i.d.
multifractal processes. It is worthwhile to note that different
techniques have been developed to simulate \cite{Wilson,Pecknold}
 or analyse \cite{Lavalee,Schmitt} multifractal
fields within the framework of Universal multifractals. Let us
emphasize that the rather straightforward "Double Trace Moment"
technique yields rather directly an estimate of the L\'evy index
$\a $ of the generator.

Furthermore, in most recent developments of multifractal studies,
in particular those related to predictability of multifractal
processes \cite{Marsan} , one needs having a  kinetic equation for
the generator, because the orientation of time axis becomes
essential, contrary to earlier simulations, where the generator
was obtained by isotropic (in time as well in space) fractional
integration over a white noise.

As there is the need for a kinetic equation in the context of
other examples/applications of L\'evy laws in physics, this paper
is devoted to establishing the corresponding "Fractional
Fokker--Planck" equation.

In this paper we consider the time evolution of a stochastic
variable forced by a random source having a stable distribution,
i.e. a generalized Langevin equation (Sect. \ref{ssec.Langevin}).
We derive the corresponding kinetic equation for the distribution
function of this stochastic variable, i.e. the Fractional
Fokker--Planck equation which has fractional space derivative
instead of the usual Laplacian (Sect.\ref{ssec.Fokker_Planck}). We
show that the expression of the Fractional  Fokker--Planck
equation is not unique (Sect.\ref{sec.nonuniqueness}) and
determine its scale invariance group, as well as its scaling
solutions (Sect.\ref{F-P Scaling}). We also generalize the
Einstein relation between the statistical exponents and the
diffusion coefficient (Sect\ref{sec.Einstein}). This helps us to
clarify the physical meaning of the exponents characterizing a
stable distribution. In Sect.\ref{sec.comparison}, we compare  our
approach to those  followed by \cite{Fogedby, Compte,
Chaves,Boland}. Finally, we discuss (Sect.\ref{applications}) the
possible applications of the theoretical results that we have
obtained.

\section{Fractional Fokker--Planck equation}
\subsection{Generalized Langevin equation} \label{ssec.Langevin}

We start with the Langevin--like equation for a stochastic
quantity $X(t)$:

\begin{equation}\label{1}
{d X(t)\over dt } = Y(t)
\end{equation}

In the classical theory \footnote{Following the approach of
Einstein and  Schmoluwski we neglect the inertial term  for large
time lags, and therefore consider the balance between viscous
friction and forcing.} of a Brownian motion, $X(t)$ is the
location of Brownian particle under the influence of stochastic
pulses $Y(t)$ \footnote{These pulses correspond to a force divided
by the friction coefficient, i.e. the inertial mass divided by the
viscous relaxation time.}. The statistical properties of this
stochastic forcing will be specified below. We first need to
derive an equation for the distribution function

\begin{equation}\label{2}
p(x, t) = \langle \d[x - X(t)]\rangle
\end{equation}

\noindent where the brackets $\langle ... \rangle $ denote
statistical averaging over stochastic force realisations. Due to
the fact that the Dirac function is the Fourier transform of the
unity, we have:

\begin{equation}\label{3}
\d [x-X(t)] =  \int_{-\infty}^\infty {d k\over 2\pi }exp \{ -i k
[x-X(t)] \}
\end{equation}

When averaged, Eq.\ref{3} yields merely that the probability is
the inverse Fourier transform of the characteristic function $Z_X
(k,t)$ (see the Appendix A for an alternative derivation
exploiting more directly this property):

\begin{equation}\label{3b}
 Z_{X} (k,t) =  \langle exp ( i k X(t)] ) \rangle
\end{equation}

\begin{equation}\label{3c}
p(x, t) =   F^{-1} [ Z_{X} (k,t)]
\end{equation}

\noindent where $F$ and $F^{-1}$ denote respectively the
Fourier--transform and its inverse:
\begin{equation}\label{fourier}
F[f]= \hat {f} (k) = \int_{-\infty}^\infty {d x} ~~  exp (i kx)
f(x) ~~~~ F^{-1} [\hat {f}]= f(x) =  \int_{-\infty}^\infty {d
k\over 2\pi }~~ exp ( -i kx) \hat {f} (k)
\end{equation}
On the other hand, Eq.\ref{1} can be integrated into:

\begin{equation}\label{3d}
 X(t) = X(0) + \int_0^t d\tau Y(\tau )
\end{equation}

\noindent Since we can assume \footnote{Indeed, we are considering
only the 'forward' Fokker-Planck equation.} without loss of
generality that $X(0) = 0$, we obtain the following equation:

\begin{equation}\label{4}
{\partial p \over \partial t} =  F^{-1} [{\partial  \over \partial
t} \left\langle exp \left[ i k \int_0^t d\tau Y(\tau
)\right]\right\rangle]
\end{equation}

Now, to make a further step, it is necessary to specify the
statistical properties of the stochastic source. We consider the
particular example \cite{Chechkin} when the source is represented
as a sum of independent stochastic "pulses"  acting at equally
spaced times $t_j$ \footnote{See Sect.\ref{sec.comparison} for an
alternative corresponding to a power-law distribution of the
waiting times \cite{Fogedby, Compte}}:

\begin{equation}\label{5}
Y(t) = \sum^\infty_{j=0} Y_{j,\Delta} \Delta \d (t-t_j) ~~~~ .
\end{equation}

\noindent where $t_0=0, t_{j+1}- t_j = \Delta$  $(j=0,1,2,... .)$
and the pulses $Y_{j,\Delta}$ are independent stochastic variables
having stable L\'evy distribution $P\{Y_{j,\Delta}\}$ for all $j$
and which has the following characteristic function \cite{Levy}

\begin{equation}\label{6}
Z_{Y_{j,\Delta}}(k) = \langle exp(ikY_{j,\Delta}) \rangle = exp
\Delta \left\{i\g k - D|k|^\a \left[ 1- i\b {k\over |k|} \om(k,
\a)\right]\right\}
\end{equation}

\noindent where $\a , \b , \g , D $ are real constants ($0<\a \leq
2, -1 \leq \b \leq 1,D\geq 0$) and $\om(k, \a)$ is defined as:

\begin{equation}\label{7}
\a \not= 1:~~ \om (k,\a ) = tan {\pi \alpha \over 2}; ~~~~~ \a =
1:~~\om (k,\a ) ={\pi \over 2} log|k|
\end{equation}

\noindent $\a $ and $\b $ classify the type of the stable
distributions up to translations and dilatations:  with given $\a
$ and $\b $,  $\g $ and $D$ can vary without changing the type of
a stable distribution. The parameter $\a $ characterizes the
asymptotic behaviour of the stable distribution:

\begin{equation}\label{7b}
p(x)\sim x^{-1-\a }, x \to \infty
\end{equation}

\noindent hence, corresponds to the critical order of moments for
their divergence:

\begin{equation}\label{7c}
\mu \ge \a : \langle x^\mu \rangle= \infty ,
\end{equation}

For (additive) walks $\a $ is also related to the fractal
dimension of the trail \cite{Mandelbrot}, whereas for the
generator of the (multiplicative) universal multifractals it
measures their multifractality \cite{Schertzer}. The parameter $\b
$ characterizes the degree of asymmetry of distribution function.
Indeed, if $\b = 0$, then negative and positive values of
$Y_{j,\Delta}$ occur with equal probabilities, while if $\b =1$ or
$ \b = -1$ (maximally asymmetric distributions) then, for $0<\a <1
$ and $ \g =0~~ P\{Y_{j,\Delta}\}$ vanishes outside from $[0,
+\infty]$ or respectively from $[-\infty , 0]$ \footnote{For $\a
>1$, $P\{Y_{j,\Delta}\}$ decays faster than an exponential on the
corresponding half axis.}. We already mentioned that maximal
asymmetry is required for generators of universal multifractals;
let us add that in this case the Laplace transform is more
convenient than the Fourier transform. The nonzero value of $\b $
implies the existence of a primary direction of the stochastic
pulses (that is, the direction to plus or minus infinity), and
thus the existence of a drift for particles in this direction. For
more details concerning the properties of stable laws see, e.g.
\cite{Gnedenko}. The meaning of $\g $ and $D$ will be discussed
and clarified below.

Now, using Eq.\ref{5} and the independence condition of the
stochastic pulses $Y_{j, \Delta}$ we get:

\begin{equation}\label{8}
\left\langle exp\left[ik\int_0^t d\tau y(\tau
)\right]\right\rangle = \left\langle exp\left[ik \sum^n_{j=0}Y_{j,
\Delta} \right]\right\rangle = \prod_{j=0}^n\langle exp\langle ik
Y_{j, \Delta}\rangle\rangle = \langle exp (ik Y_{j,
\Delta})\rangle^n
\end{equation}

\noindent where $n$ is a number of pulses corresponding to the
present time $t=n \Delta$. Therefore , with the help of the
equation of the  characteristic function of the pulses
(Eq.\ref{6}), we obtain the characteristic function $Z_X(k,t)$ (
Eq.\ref{3b}) of the stable process:

\begin{equation}\label{9}
Z_{X}(k,t)=\left\langle exp\left[ik\int_0^t d\tau Y(\tau
)\right]\right\rangle = exp \left\{ t \left[i\g k - D|k|^\a
\left(1- i\b {k\over |k|} \om (k, \a )\right)\right]\right\}
\end{equation}

The fact that this process has stationary independent increments
\cite{Skorohod} (i.e. pulses $Y_{j,\Delta}$) gives the possibility
to get directly Eq.\ref{9} without using any discretisation of
$Y(t)$ as previously done (Eq.\ref{5}). Such a derivation is
presented in Appendix A.

Now inserting this expression of  $Z_X(k,t)$ into Eq.\ref{4}, one
obtains:

\begin{equation}\label{10}
{\partial p\over \partial t}= \int_{-\infty}^\infty {dk\over 2\pi
} [i\g k - D|k|^\a + i\b D\om (k, \a )k|k|^{\a -1}] Z_X(k,t)
exp(-ikx)
\end{equation}

For the sake of the simplicity of notations, we will consider in
the following only the case $\a \not= 1, \   or \ \b=0 $.
Therefore, Eq.\ref{7} reduces to:

\begin{equation}\label{7e}
\om(k,\a )\equiv \om (\a )=tan{\pi\a \over 2}\
\end{equation}

\subsection{An expression of the Fractional Fokker--Planck Equation}
\label{ssec.Fokker_Planck} One can see that in Eq.\ref{10} the
following type of integrals appears $ F^{-1}(|k|^\a Z_X] $, which
in fact correspond to fractional differentiations. Indeed, one may
use Laplacian power for the Riesz's definition of a fractional
differentiation since for any function $ f(x)$:

\begin{equation}\label{7f}
 -\D f(x) = F^{-1}(|k|^2  \hat {f}(k))
\end{equation}

\noindent yields a rather straightforward extension:

\begin{equation}\label{14b}
 (-\D )^{\a /2}f(x) = F^{-1}(|k|^\a  \hat {f}(k))
\end{equation}

Then,  Eq.\ref{10} yields:

\begin{equation}\label{14}
{\partial p \over \partial t} + \g{\partial p \over \partial  x} =
- D \left[(-\D )^{\a /2} p + \b\om (\a ) {\partial \over \partial
x} (-\D )^{(\a -1)/2} p\right]
\end{equation}

\noindent which for symmetric laws $\b = 0$ is a straightforward
generalization of the classical Fokker--Planck equation, by:

\begin{equation}
\D \to - (-\D )^{\a /2}
\end{equation}

This also points out that the scale parameter $D$ of the L\'evy
distribution corresponds to the diffusion coefficient of the
Fractional Fokker Planck equation. On the other hand, the second
term in the left hand side of Eq.\ref{14} has an obvious physical
meaning. Independently on the value of  $\a $, it describes the
convection of particles by the (constant) velocity $\g$. For $ \a
>1,~ \g $ corresponds furthermore to the mean value of the source
$\langle Y(t)\rangle $, whereas it is no more the case for $\a
\leq 1$ since the latter is no longer finite. In the latter case,
the diffusion term has a a derivation order smaller or equal to
the convection term. This confirms that the case $\alpha =1$ is
indeed critical between two rather distinct regimes and it is more
involved than other cases. Besides, it is worthwhile to note the
role of the term (on the r.h.s.) related to asymmetry $(\b\not=0)
$. On the one hand, this term can be interpreted as an additional
contribution to the convection due to existence of the preferred
direction of the pulses related to $(\b\not=0) $. On the other
hand, such a flow is not proportional to $p$ (as the convective
flow does) but rather to $(-\D )^{(\a -1)/2} p$, which is rather
typical for the diffusion flow. In some sense, due to this term
the division of flows into convective and diffusion ones (as done
in the standard Fokker--Planck equation) becomes rather
questionable and presumably no longer relevant for the Fractional
Fokker--Planck equation. One may note that a somewhat similar
weakening of this distinction occurs also in the classical
Fokker-Plank for nonlinear systems \cite {Kampen2}. On the other
hand, it is easy to check that the Fractional Fokker--Planck
equation is Galilean invariant, as is should be: the velocity of
the  moving framework just add  to $\g$.

\subsection{The non uniqueness of the expression of the Fractional
Fokker-Planck Equation} \label{sec.nonuniqueness}

One cannot expect to obtain a unique expression for the Fractional
the Fokker-Planck equation, since there is not a unique
generalization of the differentiation to a fractional order.
Indeed, there exist various definitions of the fractional
 differentiation (see, e.g. \cite {Erdelyi}  and references therein)
which are not equivalent. This will be illustrated by two examples
in the next section. The first one is  related to the fact that
there are 'signed' (fractional) differentiation and respectively
'unsigned' (fractional) differentiations, i.e. differentiations
which  are not invariant and respectively invariant with the
mirror symmetry $x \to -x$. In the case of standard
differentiation, the question of signs is fixed:  'signed' and
'unsigned' differentiations correspond merely to odd and
respectively even orders of differentiation (hence the unique
expression of the classical Fokker--Planck equation, which is of
second order). This is no longer the case for fractional
differentiations.

The second example corresponds to the fact that fractional
differentiations are in fact defined by integration, and therefore
can depend on the bounds of integration.

Nevertheless, we are convinced that the expression corresponding
to Eq.\ref{14} is at the same time the simplest one to derive and
the one whose physical significance is the most straightforward.
On the other hand, let us emphasize that the existence of distinct
expressions for the Fractional Fokker-Planck equation does not
question the uniqueness of its solution. Indeed, these distinct
expressions are equivalent because their solution should
correspond to the unique probability density function
corresponding to a given Langevin--like equation (Eq.\ref{1}).

The non uniqueness could be rather understood in the following
way: corresponding to the distinct fractional differentiations
(and their corresponding fractional integrations), there should be
distinct ways of solving the Fractional Fokker-Planck equation in
order to obtain its unique solution.

\subsection{Two alternative expressions of the Fractional Fokker-Planck
Equation} Contrary to the unsigned fractional power of a Laplacian
Eq.\ref{14b}, let us consider for instance the following 'signed'
fractional differentiation:

\begin{equation}\label{11}
{\partial^\a \over\partial x^\a } f(x) = F^{-1}[ (-ik)^\a \hat
{f}(k)] .\end{equation}

With the help of (i) the identity ($\theta (k)$ being the
Heaviside function):

\begin{equation}\label{11b}
|k|^\a = k^\a [\theta (k) +(-1)^\a \theta (-k)]
\end{equation}

\noindent and of (ii) the inverse Fourier transform of the
Heaviside function:

\begin{equation}\label{11c}
 F^{-1} [ \theta (k)] ={1\over 2} \d (x) +{1\over  2\pi i x}
\end{equation}

\noindent as well as of (iii) the property that a Fourier
transform of a product corresponds to the convolution of the
Fourier transforms, one derives from Eq.\ref{10} an another form
of the Fractional Fokker--Planck equation.

\begin{equation}
\label{12}
{\partial p \over \partial t} +\g {\partial p \over \partial x}  =
- D \left(cos {\pi\a\over 2}+ \b sin {\pi\a\over 2} tan
{\pi\a\over 2}\right){\partial^\a p \over \partial x^\a}
-  D(1 - \b ) sin {\pi\a\over 2} {\partial^\a  \over \partial
x^\a} \int_{_{-\infty}}^{^\infty} {d x^\prime \over \pi}
{p(x^\prime, t) \over x-x^\prime}
\end{equation}

Indeed, with the help of the following determinations\footnote{One
may note that the existence of other determinations confirms the
non uniqueness of the fractional derivative defined in
eq.\ref{11b}. Furthermore, taking another determination will
merely modify some prefactors in r.h.s. of Eq.\ref{12}} $(-i)^\a =
e^{-i{\a \pi  \over  2}}, (-1)^\a = e^{-i{\a  \pi}}$, Eq.\ref{11b}
yields:

\begin{equation}\label{11d}
 |k|^\a = (-ik)^\a [\theta (k) e^{i{\a \pi \over  2}} + \theta (-k)
e^{-i{\a  \pi\over 2}}]
\end{equation}

\noindent and with the help of Eqs.\ref{11},\ref{11c},\ref{11d},
it is rather straightforward to derive Eq.\ref{12}.

However, Eq.\ref{12} is already rather involved in the case $\b
=0$, whereas this case is obvious for the equivalent Eq.\ref{14}:

\begin{equation}\label{13}
{\partial p \over \partial t} = -\g {\partial p \over \partial x}
-D cos {\pi\a\over 2} {\partial^\a p \over \partial x^\a}-D sin
{\pi\a\over 2} {\partial^\a  \over \partial x^\a}
\int_{_{-\infty}}^{^\infty} {d x^\prime \over \pi}  {p(x^\prime,
t)\over x-x^\prime}
\end{equation}

\noindent the last term of the r.h.s. of Eq.\ref{13} is rather
complex, whereas indispensable. Indeed, there is a need of signed
second term to counterbalance the first signed term of Eq.
\ref{13}, in order that the r.h.s. of Eq.\ref{13} will correspond
to an unsigned differentiation (the fractional power of the
Laplacian in  Eq.\ref{14}). Both terms correspond to the signed
fractional differentiation of order $\a$ but whereas it is applied
to  $~p~$ in the former term, it is applied to an integration of a
zero order of $~p~$ in the latter term. This zero order
integration  corresponds to the effective interaction of particles
having a scaling law inversely proportional to the distance
between them. An analogy with the interaction between dislocation
lines \cite{Landau} can be mentioned. It is plausible that the
collective effect corresponding to this the effective interaction
of particles could be responsible of the large jumps which are so
important  in L\'evy motions.

An other expression of the Fractional Fokker--Planck equation can
be also obtained with the help of the Riemann--Liouville
derivatives. The $\mu$-th order Riemann--Liouville derivatives on
the real axis are defined as

\begin {equation}
\label{15a}
({\bf D}_+^\mu f)(x) = {1 \over \G (1-\mu )} {d\over d x}
\int_{-\infty}^x d x' {f(x')\over (x-x')^\mu}~~~~~
\\ 
({\bf D}_-^\mu f)(x) = - {1 \over \G (1-\mu )} {d\over d x}
\int^{^\infty}_x d x' {f(x')\over (t-x')^\mu}
\end {equation}

\noindent where ${\bf D}_-^\mu, {\bf D}_+^\mu $ are  respectively
the left-side and the right-side derivatives of fractional order
$\mu$ ($0<\mu <1)$ and $\G $ is the Euler's gamma-function.
Appendix B gives a derivation of the corresponding expression of
the Fractional Fokker--Planck equation which is:

\begin{equation}\label{15}
{\partial p \over \partial t} + \g {\partial p \over \partial x} =
- D {\bf D}^{\a /2}_+ {\bf D}^{\a /2}_- p - D \b \om (\a )
{\partial \over \partial  x} {\bf D}^{(\a -1)/2}_+{\bf D}^{(\a
-1)/2}_- p
\end{equation}

\section{Scaling properties of the Fractional Fokker-Plank
equation}\label{F-P Scaling}

Now let us return to one of the equivalent expressions
(Eqs.\ref{14}, \ref{12}, \ref{15}) of the Fractional Fokker-Plank
equation and consider its scaling group properties which defines
fundamental properties of its solutions. Without loss of
generality we can assume $\g =0$. The scale transformation group
can be written in the following manner:

\begin{eqnarray}
& t=\l t^\prime ,~~~x = \l^\chi x^\prime , ~~~ D =\l^\kappa
D^\prime
\\ & p(x,t;\a ,\b ,D)=\l^\d p((x^\prime ,t^\prime ;\a ,\b ,D^\prime)
\end{eqnarray}

Here $\chi ,\k , \d $ are  yet unknown exponents of the scale
transformations which should leave invariant the Eqs.\ref{14},
\ref{12}, \ref{15}. One may note that the subgroup not dealing
with  D transformations was used by \cite{Fogedby} in order to
obtain the fractional order of differentiation of the Fractional
Fokker-Planck equation corresponding to a symmetric stable
processes (see Sect.\ref{sec.comparison}). Due to the further
normalization condition for the distribution function, one obtains
the following 2-parameters scale transformation group:

\begin{equation}\label{16}
 t= \l t^\prime , ~~~x = \l^\chi x^\prime , ~~~
D= \l^{\a \chi -1} D^\prime ~~~.
\end{equation}

\begin{equation}\label{17}
p(\l^\chi x,\l t; \a ,\b , \l^{\a \chi -1} D) = \l^{-\chi} p
(x,t;\a ,\b ,D)~~~.
\end{equation}

\noindent $\chi ,\l $ being the arbitrary group parameters.

If furthermore the initial condition $p(x,0)$ is invariant under
the scaling group  (Eq.\ref{16}), then the corresponding solution
of Eqs.\ref{14}, \ref{12}, \ref{15} remains invariant under the
action of this group for any other time. The simplest example of
an invariant initial condition is $ p(x,0)= \delta (x)$, which is
of fundamental importance since it corresponds to the Green
functions of Eqs.\ref{14}, \ref{12}, \ref{15}.

Let us analyze the general properties of the invariant solutions
in a similar way to renormalization group approach
\cite{Stueckelberg,Shirkov} (analogous consideration was used in a
more complex variant when calculating the spectrum of a
compressible fluid \cite{Moiseev}). Due to the fact that the
scaling invariant solutions should satisfy the identity
(Eq.\ref{17}) for any value of the arbitrary parameters $\l ,\chi
$, they could depend only on products of variables which are
independent of them. Therefore, due to the relationships $\a \chi
- \kappa -1=0$ and $\delta + \kappa = 0$, the scaling solutions
are:

\begin{equation}\label{19}
p(x,t) = {1 \over x}\Phi \left({x^\a \over Dt }\right) \equiv {1
\over (Dt)^{1 \over \a}} \Psi \left({x^\a \over {Dt}} \right)
\end{equation}


\noindent where $\Phi (.) = p(1,.) , \Psi =p(.,1) $ are arbitrary
functions which are determined by the initial conditions.
Therefore, Eq.\ref{19} represents the general form of the
invariant solutions of the kinetic equation Eqs.\ref{14},
\ref{12}, \ref{15}.


Eq. \ref{19} can be obtained by first differentiating Eq.\ref{17}
with respect to $\l $, and then, to $\chi$ and setting $\l =1,
\chi = -1/\a$. This yields the following system of equations:

\begin{equation}\label{18}
t {\partial p \over \partial t} +{x\over \a}{\partial p \over
\partial x} =- {1\over \a} p, ~~~~~~~ x {\partial p \over \partial
t} + D\a  {\partial p \over \partial D}= -p
\end{equation}

\noindent which are linear and therefore can be solved by the
method of characteristics and their solution indeed correspond to
Eq.\ref{19}.

\section{Some particular solutions}
\subsection {Explicit solutions}
With the exception of the three following cases, there is no way
to obtain an explicit expression of the solutions, with the
initial condition $f(x,0)=\d (x)$, of the Fractional Fokker-Plank
equation (Eqs.\ref{14}, \ref{12}, \ref{15} in a closed form with
the help of elementary functions:

1) $\a =2$ ($ \b =0$): \\ It corresponds to Gaussian distribution
of the stochastic forcing and to the classical Fokker--Planck
equation:

\begin{equation}\label{20}
{\partial p\over \partial t}+\g {\partial p\over \partial x} ={D}
{\partial^2 p\over \partial x^2 } ~~.
\end{equation}

\noindent which solution is the normal distribution:

\begin{equation}\label{21}
p(x,t) = {1\over \sqrt{4\pi Dt} } exp \left[-{(x-\g t)^2\over
4Dt}\right]
\end{equation}

2) $\a =1, \b =0$: \\ It corresponds to a forcing having a Cauchy
distribution and to the following Fractional Fokker-Planck
equation:

\begin{equation}\label{22}
{\partial p\over \partial t}+\g {\partial p\over \partial x} ={D}
{\partial \over \partial x}\int_{-\infty}^\infty {dx^\prime \over
\pi} {p(x^\prime , t)\over x^\prime -x } .
\end{equation}

The solution of Eq.\ref{22} with the initial condition $f(x,0) =
\d (x)$ is:

\begin{equation}\label{23}
p(x,t) = {Dt\over \pi} {1\over (x-\g t)^2 +D^2t^2} ,
\end{equation}

3) $\a =1/2,\b =1.$
\\ Then the expression  displayed in Eq.\ref{12} of the Fractional
Fokker-Planck equation takes then the form:

\begin{equation}\label{24}
{\partial p\over \partial t}+\g {\partial p\over \partial x}
=\sqrt{2} {D} {\partial^{1/2} p\over \partial x^{1/2} } .
\end{equation}

The solution of Eq.\ref{24} with the initial condition $f(x,0) =
\d (x)$ is:

\begin{equation}\label{25}
p(x,t) =\theta (x- \g t) {Dt \over \sqrt{2\pi (x- \g t)^{3/2} }}
exp \left[ -{D^2t^2 \over 2(x- \g t)}\right] ,
\end{equation}

It is easy to check that all the explicit stable distributions
(Eqs.\ref{21}, \ref{23}, \ref{25}) belong to the class of scale
invariant solutions.

\section{Generalization of Einstein relation and anomalous diffusion
coefficient} \label{sec.Einstein}

The scaling analysis of the moments of the distribution function
will lead to the generalization of the Einstein relation for the
anomalous diffusion. However, there is an important difference,
since moments of order larger than $\a <2~$ will diverge and in
particular: $\langle x^2\rangle =\infty $. On the other hand, the
motion remains mono-fractal, since all the moments $\langle
x^\mu\rangle  ~~(0<\mu <\a )$ will have the same scaling law.

The statistical moments of the distribution function of particles
initially concentrated at the origin ($p(x, 0)=\delta (x)$)
correspond to :

\begin{equation}\label{28}
\langle x^\mu\rangle = \int_{-\infty }^\infty  dxx^\mu
\int_{-\infty }^\infty {dk\over 2\pi} exp \left[-ikx -D|k|^\a
t\left(a-i\b{k\over|k|} tan {\pi\a \over 2}\right)\right] ,
 \end{equation}

In agreement with the scaling properties obtained in Sect.\ref{F-P
Scaling}), we have:

\begin{equation}\label{28a}
r_{\mu}(t)\equiv \langle x^\mu\rangle^{1/\mu} = (\a D t)^{1/\a } C
(\a ,\b ,\mu )
\end{equation}

\noindent which is obtained by renormalizing $x$ and  $k$ by $(D
t)^{1/\a }$, i.e. by considering the following variables:

\begin{equation}\label{28b}
 x_1 = k (\a D t)^{1/\a} , x_2 = {x\over (\a D t)^{1/\a}}
\end{equation}

\noindent which yield from Eq.\ref {28} the following prefactor
(depending neither on time nor on $D$) :

\begin{equation}\label{28c}
C (\a ,\b ,\mu ) = \left\{\int_{-\infty }^\infty dx_2x_2^\mu
\int_{-\infty }^\infty {d x_1\over 2\pi } exp\left[-i
x_1x_2-({|x_1|^\a \over \a} \left(1 - i\b sgn (x_1) tan({\pi\a
\over 2})\right)\right]\right\}^{1/\mu }
\end{equation}

It follows from Eq.\ref{28a} that the scaling of $r_{\mu}(t)$ in
respect to $D$ and $t$ is universal and does not depend on the
order $\mu$ of the moment considered $(\mu < \a )$. Therefore the
Einstein relation can be formulated in terms of any of the finite
moments. Indeed, only the the numerical prefactor $C (\a ,\b ,\mu
)$ (Eq.\ref{28c}) depends on $\mu$, but neither on time nor on
$D$.

The Gaussian case yields the classical Einstein formula :

\begin{equation}\label{28d}
 r_2(t) = (2D t)^{1/2}
\end{equation}

Not surprisingly, Eq.\ref{28a} confirms the fact, already pointed
out on Eq.\ref{14}, that $D$ does corresponds to a (generalized)
diffusion coefficient.

On the other hand, let us confirm that the scaling behaviour
(Eq.\ref{28a}) is independent of the initial distribution of the
particles. This is due to the fact that the distribution with
initial condition $f(x, 0)=\d (x)$ plays the role of the Green
function for the distributions with other conditions, and
therefore imposes its scaling on time and on $D$. This is easily
confirmed with the help of Eq.\ref{19} which gives the general
expression of the scaling probability densities:

\[ r_{\mu}(t) = ( D t)^{1/\a}\tilde{C} (\a ,\b ,\mu )
\]


with:
\begin{equation}\label{28f}
\tilde{C} (\a ,\b ,\mu ) =\left\{ \int_{-\infty}^\infty d\xi
\xi^{\mu -1} \Phi(\xi ^{\a})\right\}^{1 / \mu}
\end{equation}

\section{Comparison with other approaches} \label{sec.comparison}

As mentioned in Sect.\ref{Introduction}, particular cases of the
Fractional Fokker-Planck equation were obtained \cite{Fogedby,
Compte, Chaves}, and on the other hand a Langevin-type equation
has been obtained\cite{Boland} for the nonlinear Fokker-Planck
equation \cite{Plastino,Tsallis_Bukman} whose solutions maximize
the generalized q-entropy introduced by \cite{Tsallis} and exhibit
a Levy-like anomalous scaling.

A Fractional Fokker-Planck equation was obtained by \cite{Fogedby,
Compte} in the framework of the continuous time random walks
(CTRW's) model of anomalous diffusion  \cite{Klafter}. However,
this method does not involve directly a L\'evy process, but a walk
sharing some common behaviour of the latter, without being
equivalent to it. Indeed, the distribution of steps, which
corresponds to the probability distribution of the pulses in the
Langevin equation, is considered as a pure power-law. This
corresponds only to the asymptotic behaviour of the L\'evy
distribution, i.e. its tails, and therefore takes into account
only one of the L\'evy law parameters. This nevertheless allows to
establish scaling relations, therefore to determine the fractional
order of differentiation (see however a remark below), but not to
determine a precise expression of this fractional differentiation.
This is already the case for its coefficient, i.e. the
(fractional) diffusion coefficient of the Fokker-Planck equation,
since scaling reasoning does not yield a relation with the scale
parameter. Second, there is no simple way to deal with the
skewness parameter $\beta$ when considering the probability
distribution. Therefore, the corresponding non trivial term in the
Fokker-Planck (Eqs.\ref{14}, \ref{12}, \ref{15}) was not obtained
by \cite{Fogedby, Compte}. On the contrary, in our approach the
four parameters, which determine all the statistics of the pulses,
are all taken into account in an exact and rather straightforward
manner with the help of the characteristic function. On the other
hand, \cite{Fogedby, Compte} show an easy generalization to both
temporal and spatial memories is obtained by introduced a a second
(generalized) Langevin equation for the waiting times, which
introduces a  Fractional Fokker-Planck equation for the latter.
However, let us point out that the usual scaling reasoning does
not apply in a straightforward manner due to the divergence of
moments associated to the power-law of the L\'evy probability
distribution. Indeed, one cannot consider the scaling of the
variance of the distance $\langle r^{2}(s)\rangle$ traveled by a
particle after $s$ steps, because it is a divergent statistical
moment, i.e. equals to infinity, as soon as $\alpha \ne 2$. One
has to consider moments of order $\mu < \alpha$, which are not
only finite but are furthermore monoscaling (Sect.
\ref{sec.Einstein}). This last property means that the scaling of
the $\mu -th$ root of moment of order $\mu $ is independent of
this order, and therefore explains why the (mono-) scaling
reasoning works.

In a recent article \cite{Chaves} a different form of fractional
Fokker-Planck was introduced with the help of a phenomenological
and interesting modification of the classical Fick law into a
fractional Fick law, which is discussed in details. The question
of the existence of a corresponding Langevin--like equation
remains open. Some integral convergence problem imposes that the
power-law exponent of the probability distribution tails belongs
to $]1,2]$. This exponent would correspond to the L\'evy stability
index $\alpha $ the solution of this fractional Fokker-Planck was
a L\'evy motion. However, this is not exactly the case, although
it seems at first glance rather similar to it. Indeed the
corresponding characteristic function (see Eqs. 14-15 of
Ref.\cite{Chaves} involves $k^{\alpha}$ instead of $(ik)^{\alpha}$
for the characteristic function of a L\'evy motion (Eq. \ref{6}).
The difference is more obvious when considering the asymmetric
extension, either on the proposed Fokker-Planck equation (Eq. 17
of Ref.\cite{Chaves} or on the characteristic function (Eq. 19 of
Ref.\cite{Chaves}), since it does not include the non trivial
asymmetric term that we put in evidence (Eqs.\ref{14}, \ref{12},
\ref{15}). Nevertheless,  it corresponds to an interesting variant
of asymmetric diffusion, which  solutions could be rather close to
stable Levy distributions.

An essentially different generalization of the  Fokker-Planck
equation was obtained by \cite{Tsallis_Bukman,Boland} for
anomalous diffusion. Indeed, \cite{Tsallis_Bukman} showed that the
solutions of the nonlinear Fokker-Planck equation introduced by
\cite{Plastino,Tsallis_Bukman}, maximize the generalized q-entropy
\cite{Tsallis}, and correspond to a well defined L\'evy-like
anomalous diffusion, but with a finite variance and non zero
correlation. Both properties, which seem relevant and desirable
for many applications, in particular for the transport in porous
media, are not satisfied by a L\'evy motion.

Furthermore, \cite{Boland} demonstrated that the corresponding
Langevin-like equation has the particularity that the random
forces are modulated by a given power of the probability
distribution. This corresponds to a macroscopic feedback to the
microscopic kinetics, which is absent in our Langevin equation.

In comparison with these different works, we followed a rather
distinct approach since we started with a Langevin-like equation
with random forces which are  {\em exact} stable Levy processes,
which can be symmetric as well as asymmetric, and with no
limitation on the possible values of the Levy index $\alpha$. The
particular case corresponding the symmetric stable processes was
previously inferred by \cite{Fogedby, Compte, Chaves}. However, we
showed that in the more general case of asymmetric stable
processes, a new non-trivial term appears which has a rather
intermediate role between diffusion and convection (see
Sect.\ref{ssec.Fokker_Planck}). Furthermore, the use of the
characteristic function allowed us to obtain a generalization of
Einstein's formula. We also clarify the fact that different
expressions of the same Fractional Fokker-Planck equation are
obtained, depending on the type of fractional differentiation
which is used.

On the other hand, the conjecture issued by \cite{Tsallis_Bukman}
that 'further unification can be possibly achieved by considering
the generic case of a {\it nonlinear} Fokker-Planck -like equation
with {\it fractional} derivatives' should be closely examined, as
well as the fractional time evolution suggested by {Fogedby,
Compte}.

\section{Examples of applications}\label{applications}

\subsection{Generalisation of the Holtzmark distribution}
In the introduction of this paper, we recalled that the
gravitational force resulting from randomly and homogeneously
distributed point masses which acts on a given test point mass has
\cite{Holtzmark,Chandrasekar} has a stable symmetrical law with
$\a =3/2$. One may note that a similar problem arises with charged
particles and electrostatic forces. However, the distribution of
masses in the Universe is rather inhomogeneously distributed, e.g.
on a fractal set of fractal dimension \cite{Vaucouleurs}
$D_F\approx 1.2.$ (see \cite{garrido} for discussion and a
multifractal analysis). Let us extend the original result of
Holtzmark to the case of particles  distributed on a (possibly
fractal) space of dimension  $d$ and which mutually interact
according to a scaling law $1/L^r$, $L$ being  the distance
between two particles, e.g. the collective force ${\bf f}$ acting
on the randomly chosen particle of unit mass located at $\bf x$,
which results from the distribution of masses $m_k$ at points
${\bf x}^{(k)}$, has the following type \footnote{by 'type', we
mean that most of the algebraic details of the following equation
are irrelevant, only its scaling properties are relevant.} :

\begin{equation}\label{29}
{\bf f} ({\bf x})= \sum_k { m_k {{{\bf x}^{(k)}- {\bf x}} \over
|{\bf x} -{\bf x}^{(k)}|^{r+1}}}
\end{equation}

Due to the linearity of Eq.\ref{29}, the superposition $([m+m'])$
of two independent mass distributions ($[m]$ and $[m']$) yields a
force having the same probability distribution as the one  of the
sum of the two forces resulting from each of mass distribution,
i.e.:

\begin{equation}\label{29a}
{\bf f}([m]) + {\bf f}([m']) \stackrel {\rm d}{=} {\bf f}([m+m'])
\end{equation}

\noindent on the other hand, the fact that the mass is
concentrated on a fractal set of dimension $D_F$, it should scale
in the following manner with the (space) scale resolution $\l $,
i.e. the ration  $\l= {L \over l}$ of the outer scale $L$ over the
inner scale $l$ of the fractal set, in particular in the limit
$\l \to \infty$ :

\begin{equation}\label{29b}
m_{\l} = m_1 \l^{-D_F}
\end{equation}

\noindent which implies with the help of Eq.\ref{29} the following
scaling for the forces:

\begin{equation}\label{29c}
f_{\l}[m_{l}] \stackrel{\rm d} {=}  \l^{-r} f_{1}[m_{l}] \stackrel
{\rm d}{=} \bigg ({m_{l} \over m_{1}}\bigg)^{r/D_F} f_{1}[m_{l}]
\end{equation}

\noindent which together with Eq.\ref{29a} demonstrates
\cite{Feller} that $f_{\l}[m_{\l}]$ has a (symmetric) L\'evy
stable distribution, with a L\'evy index $\a = D_F/r$.

With this L\'evy index value for the random forces, and neglecting
their interrelations (which will be studied elsewhere), we may
define the random velocity of the test mass as defined by
Eq.\ref{1}, and its probability distribution by Eq.\ref{14}.

\subsection{The anomalous diffusion of a passive scalar by a
two-dimensional turbulence} Let us consider the velocity field
$v_i({\bf x})$ resulting from point-like vortices:

\begin{equation}\label{30}
v_i({\bf x})=\sum_k {\k\over 2\pi}{ \v_{ij}(x_j -x_j^{(k )})\over
|{\bf x} -{\bf x}^{(k)}|^2}
\end{equation}
where $\k $ is the intensity of the vortices, $\v_{ij}; ij=1,2 $
is the fundamental antisymmetric tensor, ${\bf x}^{k}$ is the
location of the $k^th$ vortex. Following \cite{Viecelli} we assume
that the  vortices are distributed on a fractal set of fractal
dimension $D_F$. We are therefore in the situation of the
generalization of Holtzmark distribution and indeed Eq. \ref{30}
is of the type of Eq. \ref{29} with $r=1$. Therefore, ${\bf
v}[(\k)]$ has a L\'evy probability distribution with $\a = D_F$.
one may note that this result can be obtained by using
straightforward calculations of the characteristic function in the
manner analogous to \cite{Chandrasekar}. The main distinction is
that the fractal distribution of the vortices had be taken into
account.

Therefore, the diffusion of a passive scalar in $2D$ turbulence
created by a fractal set of point--like vortices is defined by
Eq.\ref{14} with $\g=\b=0$ and $\a  = D_F$. It points out the
interest of using Fractional Fokker--Planck equation for the
analysis of the diffusion processes of particles in turbulent
media.

\subsection{Multifractal modeling}
However, as noted in the introduction, the most appealing area of
application of the Fractional Fokker--Planck equation could be for
$3D$ turbulence. Indeed it should play a key role for the
definition of the generators of {\sl dynamic} universal
multifractals \cite{fractals}. Let us first recall some basic
features of {\sl static} universal multifractals, i.e. defined
only on space. The corresponding field, e.g. the flux of energy
$F_{\l}$ at higher and higher resolution $\l = {L \over  l}$,
should respect the multiplicative property of the scale ratio,
i.e.:

\begin{equation}\label{eq.mult.epsilum}
F_{\Lambda=\l \cdot \l'}= F_{\l}\cdot T_{\l}(F_{\l'})
\end{equation}

\noindent where $T_{\l}$ is a scale contraction operator of ratio
$\l$, which in the simplest case is the isotropic self-similar
contraction ($T_{\lambda}({\bf x})=
 \lambda^{-1} {\bf x}$). Therefore $F_{\Lambda}$  might be defined with
the help of the generator $\Gamma$ of this group , more precisely
speaking by the exponentiation of the latter which satisfies the
following additive property:

\begin{equation}\label{eq.add.generator}
        \Gamma_{\Lambda=\lambda \cdot \lambda'}= \Gamma_{\lambda}
        +T_{\lambda}(\Gamma_{\lambda'})
\end{equation}

In order to satisfy the multiscaling power law:

\begin{equation}\label{multi_scaling}
\forall \lambda \in (1,\Lambda):\ \langle
F_{\lambda}^{q}\rangle\sim\lambda^{K(q)}
\end{equation}

\noindent the generator should have a logarithmic scale
divergence:

\begin{equation}
\Gamma_{\lambda} \sim \log\ \lambda
\end{equation}

\noindent this latter condition is obtained by a  convolution  of
a given the Green's function $g$ over a white-noise
$\gamma_{\lambda}$ (called the 'sub-generator'):

\begin{equation} \label{eq.generator.subgenerator}
\Gamma_{\lambda}=g \star \gamma_{\lambda}
\end{equation}

In the case of universal multifractals \cite{Schertzer} the
sub-generator is an extremely asymmetric and centered L\'evy
stable with a  L\'evy index $\alpha$ and the condition of
logarithmic of divergence (for a $D-$dimensional isotropic
process) corresponds to:

\begin{equation} \label{eq.stable.green.frc.int}
g^{\alpha}(\underline{x}) \propto
\mid\underline{x}\mid^{-\alpha.D_H}; D_H=\frac {D}{\alpha}
\end{equation}

However, in order to take into account the causality for
time-space processes \cite{Marsan}, it is rather more interesting
to consider a differentiation operator, i.e. to consider $g^{-1}$
rather than $g$, i.e.:

\begin{equation}
        g^{-1}(\underline{x},t)\star\Gamma_{\Lambda}(\underline{x},t)
                =\gamma_{\Lambda}(\underline{x},t)
\end{equation}

Furthermore, in order to take into account the difference of
scaling in space and time, the time and respectively space orders
of differentiation should be different. Therefore, the following
type of differential equation were considered:

\begin{equation}
        g^{-{\a \over D_{el}}} = {\partial \over \partial t} + (-\D )^{1-H_t}
\end{equation}

\noindent where the 'elliptical dimension of the space' $D_{el}= D
+1-H_t$ is the effective space-time dimension, $D$ is the
dimension of the  space cut and $H_t$ corresponds to the deviation
of the time scaling in comparison to the time scaling. The
Fractional  Fokker-Plank equation (Eq.\ref{14}) suggests that the
following fractional operators could be as well considered:

\begin{equation}
        g^{-{\a \over D_{el}}} = {\partial \over \partial t} + (-\D )^{1- H_t}
+\b\om (\a ) {\partial \over \partial {\bf x}} (-\D )^{(1- H_t
-1)/2}
\end{equation}

\noindent and for any value of $\b$ the evolution of the generator
has a microscopic interpretation with the help of the
corresponding Langevin equation, i.e. it points out that
Eq.\ref{eq.generator.subgenerator} corresponds to a (generalized)
path integral.

\section{Conclusions}

The original results obtained in this paper are the following:
\begin{enumerate}
\item the Fractional Fokker--Planck equation, i.e. the kinetic equation
describing anomalous diffusion in response to a  stochastic
forcing having a L\'evy stable distribution, which can be
symmetric, as well as asymmetric,
\item the a physical interpretation of
all the parameters of the L\'evy stable distributions, due to a
precise determination of the coefficients of the Fractional
Fokker--Planck equation,
\item the scale transformation group of the Fractional Fokker--Planck
equation, as well as corresponding scaling  solutions,
\item the universal dependence of the distribution function moments on the
diffusion coefficient and time,
\item some preliminary examples of applications, including a generalisation of
the Holtzmark distribution, two-dimensional diffusion of a passive
scalar and multifractal modeling of intermittent fields.
\end{enumerate}

We compare these results with particular cases obtained by
\cite{Fogedby, Compte, Chaves}, as well as their relations to a
nonlinear Fokker-Planck equation introduced by
\cite{Plastino,Tsallis_Bukman} whose solutions exhibit an
anomalous L\'evy-like diffusion \cite{Tsallis_Bukman, Boland}.

In summary, we believe that the kinetic equation obtained will be
useful for studying various physical systems with non-Gaussian
statistics.

\section{Acknowledgments}

This paper was supported by International Association under the
project INTAS-93-1194 and by the State Committee on Science and
Technologies of Ukraine, grant No 2/278.]. We  acknowledge
stimulating discussions with Academician S.V.Peletminsky and Prs.
J. Brannan, J. Duan, Yu.L.Klimontovich, M. Larcheveque and S.
Lovejoy. We thank Pr. N.G. Van Kampen for his very careful reading
of earlier versions of this paper. We are grateful to two
anonymous referees for their helpful comments and stimulating
suggestions.\\

\appendix

\section{Another  Way to Derive the Fractional Fokker-Plank
Equation.} \setcounter {equation}{0}
\renewcommand{\theequation}{\thesection.\arabic{equation}}

Here we present another approach to the derivation of Eq.\ref{9}.
Let be the probability distribution function $p(x, t)$ of $x(t)$
which obeys Eq.\ref{1} and having first characteristic function
$Z_X (k,t)$:

\begin{equation}
p(x, t) = F^{-1}[ Z_X (k,t)]
\end{equation}

Due to the linearity of the  Fourier transform and the fact that
the one considered applies only in space, not in time:

\begin {equation}
{\partial p \over \partial t} = F^{-1}\left({\partial Z_X (k,t)
\over \partial t}\right)
\end{equation}

If the pulses $Y_{dt} (t)^\prime s $ for infinitesimal time lag
$dt$ are independent and identically distributed variables for any
arbitrary time $t$, and have a second characteristic function $dt
K_Y(k)$:

\begin {equation}
Z_Y{}_{dt} (k,t)= exp [i dt K_Y (k)]
\end{equation}

\noindent then $X(t)$ has independent increments \cite{Skorohod}
and:

\begin {equation}
Z_X (k,t)=  exp [i t K_Y (k)]
\end{equation}

The demonstration of Eq. A.4 is rather straightforward, but can be
also obtained with the help of the time discretisation which we
used in order to obtain Eq.\ref{8}. \\

Inserting Eq.(A.4) into Eq.(A.2) we get

\begin {equation}
{\partial p\over \partial t} = F^{-1}[ iK_y(k)Z_X (k,t)]
\end{equation}

The particular case of L\'evy stable pulses $Y(t)$ corresponds to

\begin {equation}
K_Y(k)=\g k +i D |k|^{\a} \left[1-i\b {k \over |k|}\omega(k, \a
)\right]
\end{equation}

\noindent therefore, we immediately get Eq.\ref{10}, and we only
need to interpret $|k|^{\a} $ and $k|k|^{\a-1}$ in the physical
space, as done in Sect.2.\\

\section {Fractional Fokker--Planck Equation in
Terms of Riemann--Liouville Derivatives.} \setcounter
{equation}{0}
\renewcommand{\theequation}{\thesection.\arabic{equation}}
The $\mu$-th order Riemann--Liouville derivatives on the infinite
axis are defined as

\begin {equation}
({\bf D}_+^\mu)(x) = {1\over \Gamma (1-\mu )}{d\over d
x}\int_{-\infty} ^x d t {f(t)\over (x-t)^\mu }
\end{equation}

\begin {equation}
({\bf D}_-^\mu)(x) = {1\over \Gamma (1-\mu )}{d\over d
x}\int_x^{\infty}
 d t {f(t)\over (t-x)^\mu }
\end{equation}

\noindent where ${\bf D}_+^\mu , {\bf D}_-^\mu  $  are
respectively the left--side and the right--side derivatives of
fractional order $\mu$ ($0 < \mu <1), \Gamma $) is the Euler's
gamma--function.\\

The Fourier--transforms of the fractional derivatives (B.1), (B.2)
are the following:

\begin {equation}
F ({\bf D}_\pm^\mu f) = (\mp ik )^\mu \hat{f}(k)
\end{equation}

where $\hat{f}(k) $ is a Fourier transform of $f(x)$, and we have:

\begin {equation}
(\mp ik )^\mu = |k|^\mu exp \left(\mp {\mu \pi i\over 2} signk
\right)
\end{equation}

\noindent where $\mu $ is real and which yields:

\begin {equation}
 {\bf D}_+ ^\mu{\bf D}_- ^\mu f = F^{-1}\left[(-i k)^\mu (+i k)^\mu
 \hat{f}_k \right] = F^{-1}\left[k^{2\mu}\hat{f}_k \right] =
 F^{-1}\left[|k|^{2\mu}\hat{f}_k \right]
\end{equation}

therefore

\begin {equation}
 {\bf D}_+^{\a /2} {\bf D}_-^{\a /2}f = F^{-1}\left[|k|^\a \hat{f}_k
 \right]
\end{equation}

i.e.:

\begin {equation}
 {\bf D}_+^{\a /2} {\bf D}_-^{\a /2}f = (-\D )^{\a /2}
\end{equation}

\noindent Eq. B.7 establishes the equivalence between Eq.\ref{14}
and Eq.\ref{15}.

\section*{}


\end{document}